\documentclass[usenatbib]{mnras}

\usepackage{url, amssymb, textcomp, verbatim, graphicx, times}
\usepackage[usenames, dvipsnames]{color}

\newcommand{\adeg}[1]{{#1}$^{\circ}$}
\newcommand{\amin}[1]{{#1}$^\prime$}
\newcommand{\asec}[1]{{#1}$^{\prime\prime}$}
\newcommand{\thour}[1]{{#1}$^{\mathrm{h}}$}
\newcommand{\tmin}[1]{{#1}$^{\mathrm{m}}$}
\newcommand{\tsec}[1]{{#1}$^{\mathrm{s}}$}

\newcommand{\mjybeam}[1]{{#1}\,mJy\,beam$^{-1}$}

\newcommand{\hms}[3]{\thour{#1}\tmin{#2}\tsec{#3}}
\newcommand{\dms}[3]{\adeg{#1}\amin{#2}\asec{#3}}

\newcommand{\pbeam}[2]{\adeg{#1}$\times$\adeg{#2}}

\newcommand{\fermi}{\emph{Fermi} }

\def\aj{AJ}
\def\apj{ApJ}
\def\apjl{ApJ}
\def\apjs{ApJS}
\def\ao{Appl.~Opt.}
\def\aap{A\&A}
\def\aaps{A\&AS}
\def\baas{BAAS}
\def\mnras{MNRAS}

\title{Pulsar Candidates Toward Fermi Unassociated Sources}

\author{D. A. Frail$^1$, K.~P.~Mooley$^{2,5}$, P. Jagannathan$^{1,3}$, and H. T. Intema$^4$ \\
  $^{1}$ National Radio Astronomy Observatory, 1003 Lopezville Road, Socorro, NM 87801, USA; Email: dfrail@nrao.edu\\
  $^{2}$ Astrophysics, Department of Physics, University of Oxford, Keble Road, Oxford OX1 3RH, UK \\
  $^{3}$ Department of Astronomy, University of Cape Town, Private Bag X3, Rondebosch 7701, Republic of South Africa \\
$^{4}$Leiden Observatory, Leiden University, Niels Bohrweg 2, NL-2333CA, Leiden, The Netherlands\\
  $^{5}$ Hintze Research Fellow}

\begin{document}
\maketitle

\begin{abstract}

We report on a search for steep spectrum radio sources within the 95\% confidence error ellipses of the {\it Fermi} unassociated sources from the Large Array Telescope (LAT). Using existing catalogs and the newly released GMRT all-sky survey at 150 MHz we identify compact radio sources that are bright at MHz frequencies but faint or absent at GHz frequencies. Such steep spectrum radio sources are rare and constitute a sample of pulsar candidates, selected independently of period, dispersion measure, interstellar scattering and orbital parameters. We find point-like, steep spectrum candidates toward 11 {\it Fermi} sources.  Based on the gamma-ray/radio positional coincidence, the rarity of such radio sources, and the properties of the 3FGL sources themselves, we argue that many of these sources could be pulsars. They may have been missed by previous radio periodicity searches due to interstellar propagation effects or because they lie in an unusually tight binary. If this hypothesis is correct, then renewed gamma-ray and radio periodicity searches at the positions of the steep spectrum radio sources may reveal pulsations. 

\end{abstract}


\section{Introduction}

The \fermi mission ushered in a golden era in the study of neutron stars, rich in both observational and theoretical developments  \citep{car14}. There are currently more than 170 gamma-ray pulsars known,  and the pace of new discoveries continues unabated. In the Second LAT Pulsar gamma-ray catalog \citep[2PC;][]{aaa+13} one third are MSPs distributed isotropically on the sky, while the remainder are normal, mostly young pulsars concentrated in the galactic plane. Approximately half of the PSRs in the 2PC are {\it new} discoveries, found by searching for pulsations toward the unassociated LAT sources \citep{aaa+15}. Such searches are either blind, based on the gamma-ray photons alone \citep[e.g.][]{pgf+12}, or are guided by the initial detections and timing ephemerides from radio wavelengths \citep{rrc+11}. 

While the yield has been high, there are still many more pulsar-like candidates among the \fermi LAT unassociated sources toward which no radio pulsations have been found, especially at low galactic latitudes. \citet{ckr+12} argue that most of radio pulsars in the galactic plane that can be found by current telescopes have already been found by past large area surveys and that the lack of radio pulsations among these unassociated sources is because the radio beams are not pointed along our line of sight. There is some support for this hypothesis since the majority of blindly detected gamma-ray pulsars are not strong radio emitters. Formally, these so-called ``radio-quiet" pulsars are defined as having a phase-averaged flux density at 1.4 GHz S$_\nu<$30 $\mu$Jy. However, a pulsar could still be radio bright but its detection via a radio periodicity search could also be thwarted by the effects of scattering, dispersion, and/or absorption, either along the line of sight or in the immediate environment of the pulsar. In these cases a radio imaging approach may be useful by identifying compact, steep spectrum radio sources.

Historically, new pulsar classes such as  isolated millisecond pulsars \citep{eric80,rc79}  and globular cluster pulsars \citep{hhb85} have been initially identified this way, as well as unusual young or luminous millisecond pulsars \citep{strom87,nbf+95} have been initially identified through the radio imaging approach via their  steep spectra. Here we advocate this approach in order to identify pulsar candidates toward the unassociated sources from the Large Array Telescope (LAT) from {\it Fermi}. While radio spectra are routinely used in identifying promising blazar candidates \citep[e.g.][]{mtf15}, this is the first time, to the best of our knowledge, that the method is used for LAT pulsar candidates.

In this paper we use the recently completed GMRT 150 MHz All-Sky Radio Survey \citep[TGSS ADR;][]{int16}  to identify new  pulsar candidates toward \fermi unassociated sources.  
This paper is arranged as follows.  In \S\ref{method} we outline our search methods, while in \S\ref{results} we describe the properties the compact, steep spectrum radio sources. We end in \S\ref{discuss} summarizing the evidence that these are pulsar candidates and we outline the next steps needed to confirm.

\section{Methods}\label{method}

\subsection{The GMRT 150 MHz All-Sky Radio Survey }\label{sec:tgss}

Between April 2010 and March 2012 the Giant Metrewave Radio Telescope carried out a continuum survey at 150 MHz covering approximately  37,000~$\deg^2$, or 90\% of the sky down to a declination of \adeg{-53}. The survey achieves an median  rms noise level of \mjybeam{3.5}, with an approximate resolution of \asec{25}. The measured overall astrometric accuracy is better than \asec{1.55} in RA and DEC ($1\sigma$), while the flux density accuracy is conservatively estimated at $\sim 10$~percent. The data are publicly available on-line\footnote{\url{http://tgssadr.strw.leidenuniv.nl/}} in the form of \pbeam{5}{5} mosaicked images, image cutouts at user-specified locations, or in the form of a catalog of 0.63~Million radio sources.  For more details about the TGSS ADR we refer the reader to \citet{int16}.

\subsection{The 3FGL Unassociated Sources Sample}\label{sec:3fgl}

One third of all sources in the gamma-ray sky have no firm counterparts at other wavelengths and they are designated as ``unassociated'' sources. In the recent third Fermi Large Area Telescope (LAT) source catalog (3FGL) there are 3033 discrete sources \citep{aaa+15}.  Active galactic nuclei (AGN), mainly blazars, dominate the known sample (57\%), followed by single and binary pulsars (6\%), with supernova remnants, pulsar wind nebulae and other Galactic objects as the next most common sources class. Unassociated sources are the second-most populous source classification. There are 1010 unassociated sources in the 3FGL catalog or 33\% of the known gamma-ray sources.  839 of these are above the declination cutoff of TGSS ADR of Dec.$>-53^{\circ}$.

Using TOPCAT \citep{tay05} we searched the TGSS ADR catalog for all 150 MHz radio sources within the 95\% confidence error ellipses of all \fermi unassociated sources. Following \citet{pme+13} we adjusted the size of the major and minor axes of the 95\% error ellipses
($\theta_{maj}$, $\theta_{min}$) that are published in the 3FGL in order to reflect the true 3$\sigma$ distribution as determined empirically from the distribution of position offsets of radio and \fermi gamma-ray AGN. For the 3FGL (2FGL) this amounts to a factor of 1.148 (1.096) (F. Schnitzel, {\it priv. comm.}). 

We detected 1485 TGSS ADR sources toward 492 \fermi regions. A two point radio spectral index was derived for all 1485 sources at 150 MHz by cross-matching them with the 1.4 GHz NRAO VLA Sky Survey above Dec.$\geq-40^\circ$ \citep[NVSS;][]{ccg+98} and the 843 MHz Sydney University Molonglo Sky Survey \citep[SUMSS;][]{bls+99} for Dec.$<-40^\circ$. For the TGSS ADR sources, with neither a NVSS nor a SUMSS counterpart, we calculated a lower limit on the spectral index using based on the sensitivity limit of these surveys. An initial candidate list of 25 steep spectrum sources was identified from the original 1485 by requiring $\alpha\leq-1.50$. We further whittled down this list through a visual inspection of the TGSS ADR images and any other archival radio images at other wavelengths.  Apart from the aforementioned NVSS and SUMSS, we used radio images from the Faint Images of the Radio Sky at Twenty centimeters survey
\citep[FIRST;][]{bwh95} at 1.5 GHz, the Westerbork Northern Sky Survey  \citep[WENSS;][]{rtb+97} at 326 MHz, and the Very Large Array Low-frequency Sky Survey Redux \citep[VLSS$_r$;][]{lcv+14} at 74 MHz. In this manner we were able eliminate extended, steep-spectrum sources and flag suspicious candidates in confused regions as possible imaging artifacts. 

Our final list of 11 compact, steep-spectrum candidates is listed Table \ref{tab:steep}. For each radio source we give the best-fit TGSS ADR position (J2000), its Galactic longitude and latitude, the total flux density (S$_t$) and peak flux (S$_p$) at 150 MHz with errors, and a spectral index ($\alpha$). For those sources detected in more than two radio sky surveys we revised the original two-point spectral index ($\alpha$) with a simple least squares fit to a power law. The errors on $\alpha$ are at least 15\%. The real uncertainty is likely larger and results from the fact that the flux densities used in estimating $\alpha$ are taken from surveys whose measurements were made often decades apart (see \S\ref{discuss}).

Each radio source also has accompanying 3FGL information taken from the catalog that includes the 3FGL designation, and the measures of the significance of the gamma-ray detection (Sig.) spectral curvature (Curv.) and the variability (Var.).  Finally, for each radio source we estimate a likelihood ratio ($\Lambda$). This ratio is defined as the probability that the position offset between the radio and gamma-ray source is due to statistical error (i.e. a Rayleigh distribution) divided by the probability that the radio source is a background radio source unrelated to the \fermi source \citep{spt+15}. The numerator requires an estimate of the normalized offset of the radio source from the center of the \fermi ellipse. The denominator requires that we know the number of steep spectrum radio sources within a circle of radius equal to the angular separation between the \fermi source and the TGSS ADR source. At the completeness limit of the TGSS ADR, there are approximately 17 sources per square degree but we note in \citet{int16} that only about 10\% of these sources are steep spectrum (i.e. $\alpha<-1.59$), or 1.7 deg$^{-2}$.

\subsection{The Empty 2FGL Fields}\label{sec:2fgl}

\citet{spt+15} have produced  a catalog of compact radio sources at 5 and 9 GHz toward {\it all} of the 2FGL Fermi unassociated sources. After accounting for the newly found associations, they find that 117 of these gamma-ray sources have no compact radio candidate to the detection limit (1--2 mJy) of their survey. Following the identical method as in \S\ref{sec:3fgl} we found 129 TGSS ADR sources within the 95\% confidence interval of 42 of these 2FGL sources. There is a similar number of NVSS and SUMSS radio sources within these same regions. Suffice to say, these fields are only ``empty" in the sense that they lack compact radio sources at 5 and 9 GHz.

We identified three steep spectrum candidates using the same method as above. Two of these are already in our Table \ref{tab:steep} and have 3FGL counterparts. These are 2FGL J1827.4$-$0846 and 2FGL J1639.8$-$5145. A third candidate inside the 2FGL J0955.0$-$3949 error ellipse is a weak point source in TGSS ADR (77 mJy) but lies in fields where the SUMSS and NVSS are noisy and thus the spectral index ($\alpha=-1.6$) estimate is highly uncertain. It lies on the edge of the 95\% 2FGL error ellipse but it falls outside of the improved 3FGL error circle. For this reason, we do not consider this a viable candidate.


\section{Pulsar Candidates}\label{results}


Using the method described in \S\ref{method}, we found 11 pulsar candidates. We also found one known pulsar, PSR\,B1848+04, hitherto not claimed to be associated with any \fermi source. This is an old, normal pulsar with a period of 284 ms that lies within the \fermi unassociated source 3FGL\,J1851.2+0423. In this section we give a brief description about the properties of each candidate and explore the possible multi-wavelength counterparts of these candidates.

There are a number of optical-infrared associations for the majority of the candidates, but given their current positional accuracy of a few arcseconds and the proximity to the Galactic plane, the high optical source densities in these regions make it difficult to identify counterparts with certainty. X-ray observations have been undertaken by the {\it Swift} satellite toward seven of our 3FGL regions\footnote{The results are available on-line at \url{http://www.swift.psu.edu/unassociated/}} \citep{sf13}. The integration times varied from 0.1 ks to 7 ks. We also additionally checked the latest all-sky catalogs of the ROSAT, Chandra, and XMM-Newton telescopes. There are no detections of X-ray sources coincident with any our of radio candidates.

\subsection{3FGL\,J0258.9+0552}

\begin{figure}
\includegraphics[width=5in]{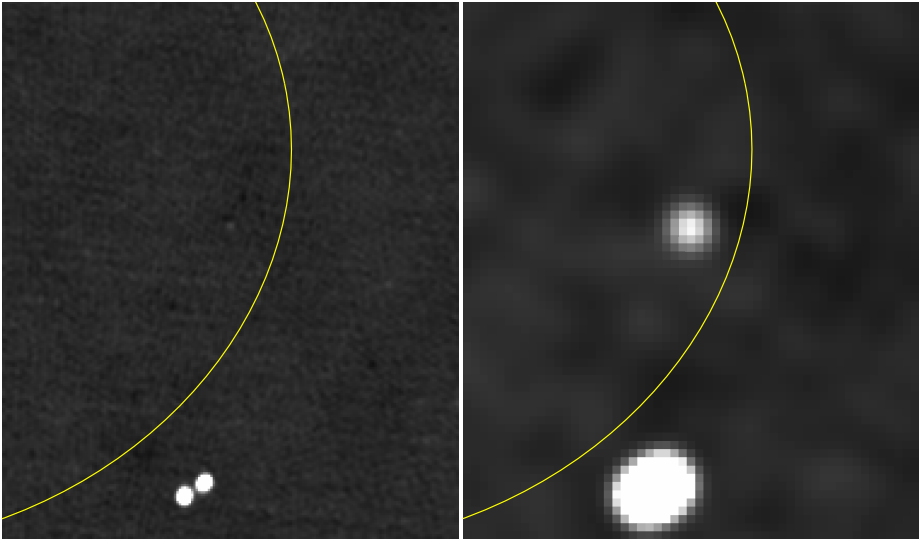}
\caption{(Right) A steep spectrum pulsar candidate toward 3FGL J0258+0552 with a low likelihood ratio ($\Lambda$) in the center of the TGSS ADR image at 150 MHz (135 mJy) shown here with the original \fermi error ellipse (yellow line). On the left is a FIRST image at 1.4 GHz showing a faint (1.2 mJy) and unresolved point source. The field of view is \amin{6.7}$\times$\amin{5.6} and the rms noise in the FIRST and TGSS ADR images are \mjybeam{0.15} and \mjybeam{3.6}, respectively.}
\label{fig:faint}
\end{figure}

This candidate has a peak flux density of \mjybeam{$129.6\pm4.3$} at 150 MHz, and the
spectral index between 150 MHz and 1.4 GHz is $-2.05$.
There is a weak detection of this source in  FIRST and NVSS images.
The former allows us to improve the position and angular resolution.
The FIRST source is unresolved at 1.4 GHz with a synthesized beam of \asec{5}, $S_t= 1.21\pm0.22$ mJy, and $S_p=1.45\pm0.14$ mJy beam$^{-1}$.
This allows precise localization of the candidate, RA=\hms{02}{58}{38.56}, DEC=\dms{05}{51}{52.8}, with uncertainties of \asec{0.2} in both coordinates.
At this location, there are no viable multiwavelength counterparts in the Vizier database. We show the FIRST and TGSS ADR images in Fig. \ref{fig:faint}.



\subsection{3FGL\,J1533.8-5231}

This candidate has a strong detection in TGSS ADR ($S_t=157.5\pm12.8$ mJy), but only an upper limit in the SUMSS catalog.
At the location of the TGSS source, the peak pixel value in the SUMSS is 1.6 mJy, implying that the spectrum is slightly steeper than that given in the Table~\ref{tab:steep}, $-1.89$.
There are no multiwavelength counterparts within \asec{5} of the TGSS ADR source in the Vizier database.

\subsection{3FGL\,J1639.4-5146}

\begin{figure}
\includegraphics[width=5in]{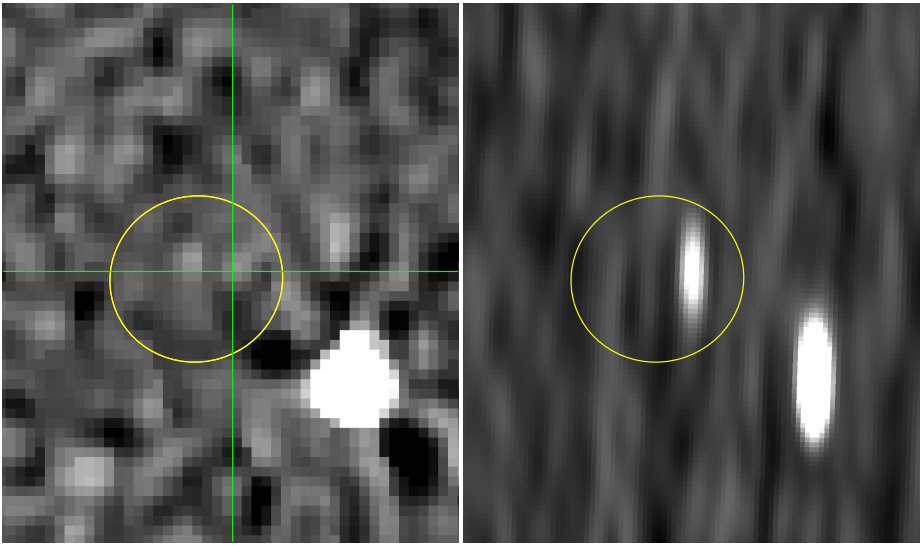}
\caption{(Right) A steep spectrum pulsar candidate toward 3FGL J1639.4$-$5146 with a high likelihood ratio ($\Lambda$) in the center of the TGSS ADR image at 150 MHz (175 mJy) shown here with the original \fermi error ellipse (yellow line). On the left is a SUMSS image at 843 MHz with (green) crosshairs indicating a faint, uncatalogued radio source ($<$3 mJy) at the same position as the 150 MHz source. The field of view is \amin{11.9}$\times$\amin{9.2} and the rms noise in the SUMSS and TGSS ADR images are \mjybeam{1.3} and \mjybeam{9.8}, respectively.}
\label{fig:bright}
\end{figure}

The 3FGL error circle is small ($\theta_{maj}$=\amin{2}) and it has a bright ($S_t=174.9\pm16.3$ mJy), compact TGSS ADR source within it. 
There is no cataloged SUMSS counterpart for the TGSS ADR source, but at that location the peak pixel value in SUMSS is \mjybeam{2.9}.
Thus, the spectral index in Table~\ref{tab:steep}, $-1.95$, is an upper limit.
A possible optical counterpart with a B band magnitude of 18.8 is catalogued in USNO-B1.0, and is located \asec{2.1} from the TGSS ADR source. We show the SUMMS and TGSS ADR images in Fig. \ref{fig:bright}.

\subsection{3FGL\,J1747.0-3506}

A point-like TGSS ADR source with $S_t=78.7\pm15.6$ mJy lies in the center of the 3FGL error ellipse ($\theta_{maj}$=\amin{6}). There are no NVSS or SUMSS counterparts at this location at the level of $>$\mjybeam{1}. 
Based on this upper limit, we derive an upper bound of $-1.77$ to the spectral index between 150 MHz and 1.4 GHz.
We note that there is some extended emission just NW, offset by \amin{1}, from this location in NVSS, which is likely unrelated to the TGSS ADR source.
There are various possible optical counterparts having V band magnitudes of $\sim$20 located within \asec{2.5} catalogued in Vizier database.

\subsection{3FGL\,J1827.6-0846}

This candidate is on the edge of the 3FGL error circle, and has a total flux density of $229.9\pm14.8$ mJy in TGSS ADR.
There is no NVSS emission at this location at a level of $<$\mjybeam{0.8}, but there is a 777$\pm$126 mJy VLSS$_r$ source.
This confirms the steep spectrum nature of the TGSS ADR source.
The spectral index between 150 MHz and 1.4 GHz is therefore less than $-2.0$, while between 74 MHz and 150 MHz it is approximately $-2.6$.
There are two possible counterparts in UKIDSS located $\sim$\asec{1.3} from the TGSS ADR location. 

\subsection{3FGL\,J1830.8-3136}

The density of the TGSS ADR source is $79.6\pm11.6$ mJy.
At this position, the NVSS registers a weak source, possibly extended with $S_t = 2.6\pm0.5$ mJy and $S_p = $\mjybeam{$0.96\pm0.55$}. 
Inspection of the NVSS image cutout suggests that this could be a sidelobe of a brighter, 54 mJy, source located to the SW.
There is only an upper limit with SUMSS of $<$\mjybeam{3.5}.
The spectral index between 150 MHz and 1.4 GHz derived assuming 2.6 mJy flux density in NVSS is $-1.53$.
There are no other counterparts within \asec{5} of the TGSS ADR source in the Vizier database.


\subsection{3FGL\,1901.5$-$0126}

This candidate is detected (strongly, with $S_t = 361.8\pm14.7$ mJy) in the TGSS ADR, but not in any of the other large scale surveys. 
The noise in the VLSS$_r$ image is very high, $\sim$\mjybeam{500}, 
so its absence in this survey is not very surprising.
Based on the flux density upper limit from NVSS, we calculate a spectral index of less than $-2.23$ between 150 MHz and 1.4 GHz.
Hence, we consider this source to be a strong candidate.
A UKIDSS source with a K band magnitude of 17.4 is located \asec{0.8} from the TGSS ADR location, and is a possible counterpart of the radio source.

%

\subsection{3FGL\,J1925.4+1727}

This candidate lies along the Galactic plane, towards the Sagittarius and Cygnus arms. 
It appears to be partially resolved in the TGSS ADR, with $S_t = 74.7\pm10.4$ mJy.
The non-detection in NVSS implies a spectral index of less than $-1.52$ between 150 MHz and 1.4 GHz.
A resolved source at 327 MHz is also reported at this position with a flux density of 71$\pm$8 mJy \citep{tgc+96}. 
This candidate may be a flat spectrum object at MHz frequencies and thus consider this source to be a weak candidate.
There is a stellar source with a R band magnitude of 17.6 located within \asec{2.5}  from the TGSS ADR location, and could be the optical counterpart of the radio source.

\subsection{3FGL\,J1949.3+2433}

This source has a total flux density of $160.4\pm11.4$ mJy in the TGSS ADR at 150 MHz and appears to be partially resolved. 
It is also detected in the NVSS at 1.4 GHz (5.1$\pm$0.5 mJy) and the WSRT Galactic plane survey at 327 MHz (31$\pm$4 mJy), but not in the VLSS$_r$. 
The 75 MHz image cutout from the latter survey is rather noisy, which appears to be the reason for the non-detection.
A fit to the flux densities from these different catalogs gives a spectral index of $-1.51\pm0.15$.
Considering the modest steepness of the spectrum together with the partially resolved nature of the source in the TGSS ADR, we claim that this source is a weak candidate.
A bright near-infrared source catalogued in 2MASS lies \asec{1.2} offset from the TGSS ADR source. It has a J band magnitude of 16.2, and is a possible counterpart of the radio source.

\subsection{3FGL\,J2028.5+4040c}

This candidate lies near Cygnus OB2 star-forming region. 
The total flux density in the TGSS ADR is $389.7\pm20.2$. The source is not seen in the  VLSS$_r$ image cutout, but the 74 MHz data is rather noisy in this Galactic plane direction. A point source is detected in the NVSS (12.8$\pm$0.7 mJy), the WENSS (218$\pm$9.4 mJy) and the WSRT Galactic plane survey (126$\pm$7 mJy). \citet{gbhw03} report 144$\pm$7 mJy and 62$\pm$9 mJy at 350 MHz and 1.4 GHz, respectively, while \citet{whl91} measure a 1.4 GHz flux density of 64$\pm$3 mJy. It appears likely that this is a variable radio source.

The steep spectrum calculated between TGSS ADR and NVSS, $-1.57$, could be a variability artifact, due to the fact that these measurements are taken at different epochs. In this case, the gamma-rays could originate from a variable AGN seen through the galactic plane \citep[e.g.][]{tgd+95}. Alternatively, this could be a neutron star in a Be binary system such as LSI+61$^\circ$303, PSR B1259$-$63 or PSR J2032+4127 \citep{lsk+15}. 

There is a stellar source catalogued in WISE, 2MASS, and SDSS \asec{5.2} offset from the TGSS ADR source. From the optical and infrared photometry we estimate its distance to be $\sim$100 pc, and the spectral type to be M, assuming that it is on the main sequence. Another optical source in the vicinity is a galaxy in SDSS that is \asec{3.6} offset from the radio source. We reject both these sources as possible counterparts to the TGSS ADR source, given that their offset distances are much larger than the uncertainty in the TGSS ADR source location.


\subsection{3FGL\,2210.1+5925}

The TGSS ADR source ($S_t=158.7\pm5.0$) lies just south of an extended, cataloged NVSS source (3.1$\pm$0.5 mJy), located \asec{4.5} away. At the TGSS ADR source position, the NVSS peak flux density is \mjybeam{2.5}. There is a WENSS 327 MHz source (57$\pm$5.1 mJy) and weak VLSS$_r$ emission at this location (\mjybeam{360$\pm$150}). A fit using all of these detections gives a spectral index of $-1.71\pm0.17$. 
There are no other counterparts within \asec{3} of the TGSS ADR source in the Vizier database.

\section{Discussion and Conclusions}\label{discuss}

Pulsars are ultra-steep spectrum radio sources with power-law spectra extending (typically) from 100 MHz to several GHz, with slopes $\alpha=-1.8\pm0.2$ \citep[where S$_\nu\propto\nu^\alpha$;][]{mkk00}.  While the {\it intrinsic} spectral indices may be slightly flatter \citep[$-1.4\pm{1.0}$;][]{blv13}, they are still considerably steeper than the canonical $\alpha=-0.7$ slope of the extragalactic sources that dominate source counts of the radio sky. In \citet{ki08}, fewer than 0.4\% of the radio sources  have $\alpha< -1.8$, or less than 1 source per 140  deg$^2$. At the steep spectrum completeness limit of the TGSS ADR we estimate that
there are 1.7 steep spectrum sources per deg$^{2}$ \citep{int16}. The only other known discrete radio class with similar spectral slopes are the luminous high redshift galaxies (HzRGs). These rare, massive, star forming galaxies are highly sought after in their own right as they are known to be excellent tracers of protoclusters in the early universe \citep{md08}. 

HzRGs are not known to be gamma-ray emitters. Thus, if our steep spectrum sources are pulsars, then the radio and gamma-ray properties of the sample should resemble that of the known \fermi pulsars. Table \ref{tab:steep} lists some of these properties taken directly from the 3FGL catalog. In their galactic distribution and their radio flux densities these candidates are similar to the known pulsars. By and large the gamma-ray properties of our candidates also resemble the known sample. Apart from the mode-changing pulsar PSR\,J2021+4026 \citep{sah+14}, transitional pulsars such as PSR\,J1023+0038 \citep{abb+15}, strongly glitching pulsars \citep[e.g.][]{pga+12}, and the Crab pulsar, gamma-ray light curves do not vary.  None of the candidates are variable gamma-ray sources; the variability indices in Table \ref{tab:steep} are all within the range of the known \fermi pulsars \citep{aaa+13}. Likewise, pulsars are known to have power-law spectra in gamma-rays with exponential cutoffs around a few GeV. Here we find some differences between known pulsars and our candidates. The spectral curvature values in Table \ref{tab:steep} are distributed on the low end of the known pulsar distribution. All but two are listed in the 3FGL as having simple power-law fits while the two candidates with the largest values of curvature are fit with Log parabolic forms. However, with some exceptions, our \fermi unassociated sources are not bright in gamma-rays (Sig.$<$10) and thus measuring significant exponential cutoffs  would be difficult in these cases. \citet{ckr+15} have cautioned against eliminating 3FGL pulsar candidates on the basis of spectral curvature when the source is not bright in gamma-rays or the diffuse background is high.

More information is required to determine whether these compact TGSS ADR radio sources within the unassociated error ellipses  are pulsars. Our positions at 150 MHz are good to only $\pm 2^{\prime\prime}$. This is sufficient accuracy to look for X-ray and (variable) optical counterparts, and to do a blind search for pulsed gamma-ray emission from normal pulsars of the brighter candidates.  However, in order 
 to reduce the computation costs for blind searches of MSPs, we need the sub-arcsecond positions that a radio interferometer provides. Such follow-up observations have the added benefit of allowing us to reject candidates. Pulsars are unresolved while HzRGs have kpc-size extended structure that will be visible at arcsecond resolution. Likewise, the radio surveys we used for calculating the spectral indices were taken two decades ago and thus some variable sources may have been falsely identified as steep spectrum.
 
The current positions of the candidates are sufficient to carry out radio pulsation searches with single dishes, but we must first understand why they have been missed in the first place. Several of our candidates are at high galactic latitudes where existing radio searches have been concentrated.  \citet{ckr+15} have noted that there remain many 3FGL unassociated sources at high galactic latitudes with strong PSR-like characteristics but for which no radio pulsations have yet been found. Radio quiet MSPs are thought to be rare so they explain this anomaly as originating from flux modulation due to interstellar scintillation, or a rapidly changing period due to the PSR being in a tight binary. Our imaging method identifies PSR candidates without regard to binary orbits, and since the 150 MHz measurements are averaged over a night's observing they can mitigate the effects of interstellar modulation somewhat. It may be that a repeated search of these 3FGL error ellipses may yield new MSPs. We note that none of our 3FGL sources are in the lists of the 3FGL sources searched at Arecibo or Parkes \citep{ckr+15,cck+16}. The situation at low galactic latitudes is less promising. The most likely explanation for the lack of detection, if these are indeed pulsars, is that the temporal broadening along the line of sight is too large. For the brighter candidates in our sample, it could be worthwhile to search at frequencies above 3 GHz where such temporal scattering effects are much reduced ($\propto\nu^{-4}$).

Summarizing, we have identified 11 compact, steep spectrum radio sources that are positionally coincident with \fermi 3FGL unassociated sources. We argue that these are strong pulsar candidates and we outline the necessary next steps that need to be taken to confirm or refute this hypothesis.

\section*{Acknowledgments}
This research has made use of data and/or software provided by the High Energy Astrophysics Science Archive Research Center (HEASARC), which is a service of the Astrophysics Science Division at NASA/GSFC and the High Energy Astrophysics Division of the Smithsonian Astrophysical Observatory.
This work has also made extensive use of the SIMBAD and Vizier databases maintained by the Centre de Donn\'ees astronomiques de Strasbourg.
KPM acknowledges funding from the Hintze Foundation. DAF thanks S. Kulkarni and T. Readhead for their hospitality at Caltech while this work was being written up. HTI acknowledges financial support through the NL-SKA roadmap project funded by the NWO.

{\it Facilities:} GMRT, Fermi (LAT).

\clearpage

\clearpage


\begin{table*}
\caption{Steep Spectrum Radio Sources Toward Fermi Unassociated Sources}
\resizebox{\textwidth}{!}{%
\begin{tabular}{ccccccllllr}
\hline\hline
RA & DEC & $l, b$ & S$_t$ & S$_p$ & $\alpha$ & 3FGL Name & Sig. & Curv. & Var. & $\Lambda$\\
\hline
02:58:38.62 ($\pm$0.02) & +05:51:51.0 ($\pm$0.4) & 170.7,$-$44.9 & 135.3$\pm$7.3 & 129.6$\pm$4.3 & $-$2.05 & J0258.9+0552 & 8.0 & 1.72 & 38.9 & 740\\
15:33:55.36 ($\pm$0.04) & -52:32:58.3 ($\pm$3.3) & 326.3,+2.8 & 157.5$\pm$12.8 & 133.2$\pm$6.5 & $-$1.89 & J1533.8-5231 & 6.4 & 1.80 & 50.0 & 21,300\\
16:39:23.83 ($\pm$0.03) & -51:46:34.1 ($\pm$2.8) & 334.2,$-$3.3 & 174.9$\pm$16.3 & 169.3$\pm$7.7 & $-$1.95 & J1639.4-5146 & 20.1 & 1.85 & 58.0 & 43,900\\
17:47:00.24 ($\pm$0.06) & -35:05:54.2 ($\pm$1.9) & 354.9,$-$3.4 & 78.7$\pm$15.6 & 90.0$\pm$8.3 & $-$1.77 & J1747.0-3506 & 11.7 & 1.85 & 64.5 & 41,300\\
18:27:36.25 ($\pm$0.03) & -08:49:41.3 ($\pm$0.6) & 22.4,+1.2 & 229.9$\pm$13.8 & 196.2$\pm$8.3 & $-$2.02 & J1827.6-0846 &  8.7 & 1.42 & 51.6 & 1700\\
18:30:38.71 ($\pm0.11$) & $-$31:35:03.9 ($\pm 3.2$) & 2.4,$-$9.8 & 79.6$\pm$11.6 & 56.9$\pm$7.3 & $-$1.53 & J1830.8-3136 & 5.3 & 2.42 & 40.9 & 1700\\ 
19:01:33.72 ($\pm$0.02) & -01:25:28.0 ($\pm$0.3) & 32.8,$-$2.9 & 361.8$\pm$14.7 & 329.2$\pm$8.8 & $-$2.23 & J1901.5-0126 & 14.0 & 1.59 & 60.4 & 22,000\\
19:25:33.03 ($\pm$0.14) & +17:22:36.1 ($\pm$1.6) & 52.2,+0.6 & 74.7$\pm$10.4 & 49.7$\pm$6.8 & $-$1.52 & J1925.4+1727 &    9.6 & 6.29 & 47.3 & 400\\
19:49:04.82 ($\pm 0.07$) & +24:32:34.4 ($\pm 1.1$) & 61.2,$-$0.7 & 160.4$\pm$11.4 & 99.5$\pm$7.6 & $-$1.51 & J1949.3+2433 & 8.2 & 5.00 & 42.2 & 1600\\ 
20:28:24.33 ($\pm 0.05$) & +40:37:49.8 ($\pm 0.6$) & 79.1,+1.1 & 389.7$\pm$20.2 & 292.6$\pm$12.6 & $-$1.57 & J2028.5+4040c & 5.0 & 3.71 & 48.8 & 5500\\ 
22:11:08.27 ($\pm$0.05) & +59:25:43.9 ($\pm 0.3$) & 103.8,+2.6 & 158.7$\pm$5.0 & 121.9$\pm$3.2 & $-$1.71 & J2210.1+5925 & 9.7 & 2.46 & 54.9 & 320\\ 
\hline
\multicolumn{10}{p{6.5in}}{Notes: (a) RA and DEC quote Gaussian fit errors only. An extra uncertainty of \asec{1.55} should be added in quadrature to these fit errors  in order to obtain the true (1$\sigma$) astrometric error for the TGSS ADR survey.}\label{tab:steep}
\end{tabular}}
\end{table*}

\end{document}